\documentclass[aps,prl,twocolumn,superscriptaddress]{revtex4-2} 
\usepackage{amsmath,amssymb,bm}
\usepackage[T1]{fontenc}
\usepackage{amsmath,graphicx,amssymb,epsfig,babel,dsfont}
\usepackage{mathrsfs}
\usepackage{color}
\usepackage{amsbsy}
\usepackage{float}
\usepackage{braket}
\newcommand{\rd}{{\rm d}}

\usepackage{xcolor}
\begin{document}

\title{
Radiative spin caloritronics in nonreciprocal many-body systems
}

\author{Philippe Ben-Abdallah}
\email{pba@institutoptique.fr}
\affiliation{
Laboratoire Charles Fabry, UMR 8501, Institut d'Optique, CNRS,
Universit\'e Paris-Saclay,
2 Avenue Augustin Fresnel, 91127 Palaiseau Cedex, France
}

\date{\today}

\begin{abstract}
We predict the spin thermal Hall effect in nonreciprocal magneto-optical many-body systems, in which a longitudinal radiative heat current generates a transverse accumulation of the spin angular momentum carried by thermal photons. We show that this effect and the inverse spin thermal Hall effect constitute an Onsager-Casimir reciprocal pair, thereby establishing heat and photon spin as coupled transport channels in nonreciprocal photonic systems. The second law of thermodynamics imposes fundamental bounds on the spin-heat coupling, leading to a thermal-spin figure of merit that quantifies the efficiency of radiative spin-heat conversion. Our results establish a complete thermodynamic framework for photon spin caloritronics and lay the conceptual foundations for spin-controlled thermal radiation and nonreciprocal photonic thermal devices.
\end{abstract}

\maketitle

The transport of heat by thermal radiation in many-body systems has recently emerged as a fertile platform for exploring transport phenomena traditionally associated with condensed-matter physics, including thermal Hall effects, nonreciprocal energy transport, and topological heat flow~\cite{pbaPRL2011,pbaRMP,pba2016,Fan,Cuevas,Cuevas2}. In magneto-optical systems~\cite{Kimel2022,Sato2022}, an external magnetic field breaks Lorentz reciprocity and couples the orbital and spin degrees of freedom of thermal photons, giving rise to fundamentally new mechanisms for controlling radiative heat transfer at the nanoscale.
A major advance in this direction was the recent prediction of the inverse spin thermal Hall effect (ISTHE) in magneto-optical many-body systems~\cite{pbaPRL2025}, in which a longitudinal gradient of photon spin angular momentum generates a transverse radiative heat flux. This discovery established photon spin as a genuine thermodynamic transport variable capable of driving energy transport through nonreciprocal electromagnetic interactions.
From the standpoint of irreversible thermodynamics, however, the inverse effect naturally raises the question of its reciprocal counterpart. If a longitudinal spin-angular-momentum gradient can induce a transverse heat flux, should a longitudinal radiative heat flux conversely generate a transverse spin-angular-momentum accumulation, or equivalently a transverse spin-angular-momentum gradient? More fundamentally, are these two conversion mechanisms connected by the Onsager-Casimir reciprocity relations despite the explicit breaking of time-reversal symmetry by the external magnetic field?

In this Letter, we answer these questions by introducing the spin thermal Hall effect (STHE) in magneto-optical many-body systems, whereby a longitudinal
radiative heat current generates a transverse photon-spin
accumulation. We show that this phenomenon is the
Onsager-Casimir reciprocal counterpart of the inverse spin
thermal Hall effect and derive the corresponding reciprocity
relations directly from fluctuational electrodynamics.Together, these results establish a unified thermodynamic framework for spin-resolved radiative heat transport and complete the thermodynamic description of spin Hall transport in nonreciprocal photonic many-body systems.

The essential conceptual point is that Onsager--Casimir reciprocity is a symmetry constraint, not a microscopic derivation of the STHE. Our previous work~\cite{pbaPRL2025} established spin-to-heat conversion, whereas the present work independently demonstrates the converse microscopic process: a purely thermal driving generates a transverse photon-spin accumulation. The latter requires the calculation of spin-resolved correlations of the thermally fluctuating electromagnetic field and is not obtained by simply reversing the ISTHE calculation. Taken together, the two effects establish the bidirectional conversion heat-photon\ spin that underlies radiative spin caloritronics.

\begin{figure} \centering 
\includegraphics[height=0.25\textwidth,angle=0]{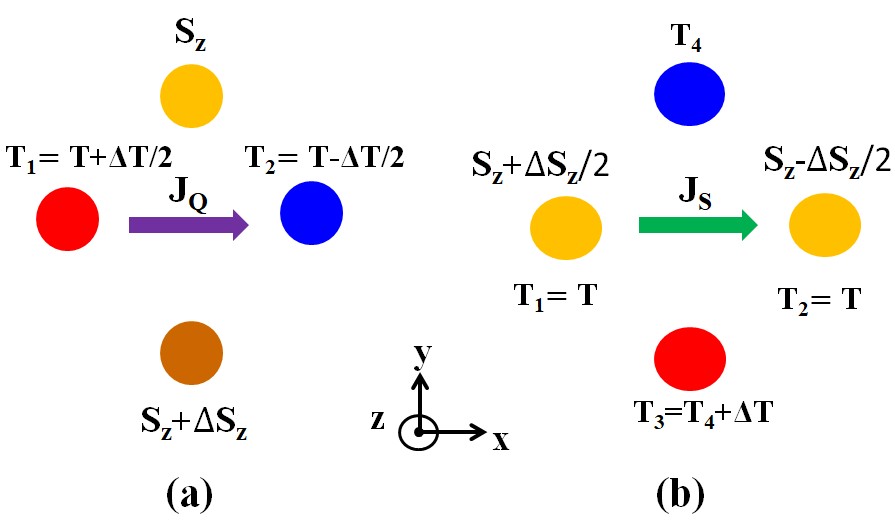} 
\caption{
Schematic of the:  (a) radiative Spin Thermal Hall Effect where  a temperature gradient $\Delta T$ and therfefore a heat flux $J_Q$ through a non-reciprocal network of C4 symmetry made with  magneto-optical nanoparticles  under the action of an external magnetic field $\bold{H}_{ext}$ in the the $z$ direction induces a variation $\Delta\bold{S}_z$ of spin angular momentum in the transversal direction of primary gradient ; (b) Inverse Spin Thermal Hall Effect where a variation $\Delta\bold{S}_z$  in one direction (due to a spatial variation of ecternal field),  and therefore a spin current $J_S$ in the same direction, gives rise to a temperture gradient $\Delta T$ in the transverse direction even without temperature gradient (i.e. $T_1=T_2=T$).
}
\label{Fig1: STHE and ISTHE}
\end{figure}

We consider a nonreciprocal many-body system consisting of four identical
InSb nanoparticles of radius $R=50\,\mathrm{nm}$ immersed in a thermal
bath at temperature $T_b=300\,K$. The particles are much smaller than the thermal
wavelength, so that their electromagnetic response can be described within
the electric-dipole approximation. They are located at the vertices of a
regular square of diagonal length $a=4R$, as shown in
Fig.~1. This $C_4$ configuration is chosen as a minimal demonstrator because its symmetry cleanly separates the longitudinal thermal driving from the transverse spin response and allows a direct comparison with the ISTHE in the same electromagnetic system. The STHE is not restricted to this geometry or to InSb: the formulation in terms of the many-body electromagnetic Green tensor applies more generally to nonreciprocal particle networks and continuous structures.
A stationary radiative heat flux is established through the network by connecting two particles located at opposite corners of the square to  thermal reservoirs maintained at different temperatures. Without loss of generality, particles 1 and 3 are held at the temperatures
\begin{equation}
T_1=T_H,\qquad
T_2=T_C,
\qquad
T_H>T_C,
\label{eq:temperature_bias}
\end{equation}
whereas particles 3 and 4 are not thermostated and reach steady-state temperatures through their radiative interactions with the other particles and the surrounding thermal bath. The thermal reservoirs continuously inject and extract the power required to maintain the prescribed temperatures, thereby establishing a stationary longitudinal radiative heat current flowing along the diagonal joining particles 1 and 2.
At steady state, the external powers supplied by the reservoirs are therefore
\begin{equation}
P_i^{\rm ext}=-\varphi_i,
\qquad
i=1,2,
\end{equation}
where $\varphi_i$ denotes the net radiative power received by particle
$i$. Since particles 3 and 4 are not connected to external reservoirs,
their steady-state temperatures are determined by the conditions
\begin{equation}
\varphi_3=\varphi_4=0.
\end{equation}
\begin{figure}\centering
\includegraphics[angle=0,scale=0.4]{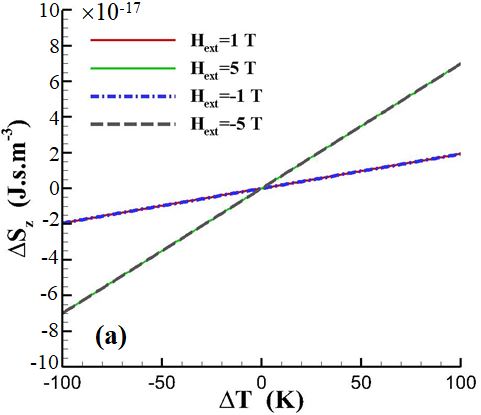}
\includegraphics[angle=0,scale=0.4]{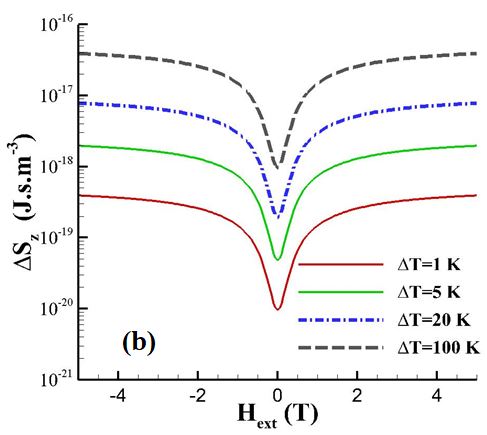}
\includegraphics[angle=0,scale=0.4]{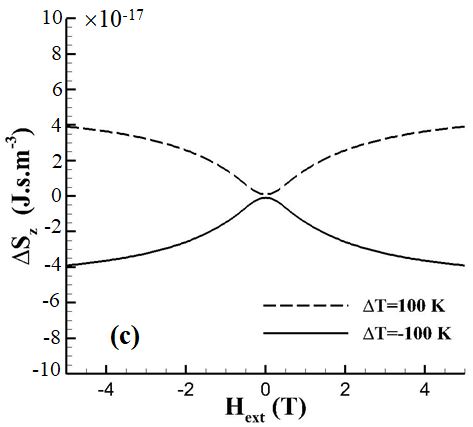}
\caption{
Thermal spin accumulation in the four-particle network of InSb nanoparticles as a function of (a) the longitudinal temperature difference $\Delta T$ for different external magnetic fields applied along the $+z$ direction; (b) the external magnetic field magnitude $H_{\rm ext}$ for different temperature differences; and (c) the external magnetic field magnitude $H_{\rm ext}$ for temperature differences of opposite sign. The optical response of the InSb nanoparticles is described by the gyrotropic Drude--Lorentz model~\cite{SupplMat}.
}
\label{Fig2: symmetry}
\end{figure}

In the presence of an external magnetic field
$\mathbf H_{\rm ext}=H_{\rm ext}\hat{\mathbf z}$,
the dielectric permittivity of a magneto-optical particle becomes gyrotropic
~\cite{Palik},
\begin{equation}
\boldsymbol{\varepsilon}(\omega,\mathbf H_{\rm ext})
=
\begin{pmatrix}
\varepsilon_{1} & -i\varepsilon_{2} & 0\\
i\varepsilon_{2} & \varepsilon_{1} & 0\\
0&0&\varepsilon_{3}
\end{pmatrix}.
\label{eq:epsilon}
\end{equation}
Here,  the off-diagonal component satisfies
$\varepsilon_{2}(-\mathbf H_{\rm ext})=-\varepsilon_{2}(\mathbf H_{\rm ext})$
and breaks Lorentz reciprocity. For spherical nanoparticles, the radiatively corrected polarizability is~\cite{Albaladejo}
\begin{equation}
\boldsymbol{\alpha}=
\left(
\mathds{1}
-i\frac{k_0^3}{6\pi}\boldsymbol{\alpha}_{0}
\right)^{-1}
\boldsymbol{\alpha}_{0},
\end{equation}
where
\begin{equation}
\boldsymbol{\alpha}_{0}
=
4\pi R^3
(\boldsymbol{\varepsilon}-\mathds{1})
(\boldsymbol{\varepsilon}+2\mathds{1})^{-1}.
\end{equation}
following the generalized Clausius-Mossotti relation for anisotropic
particles.
The polarizability can therefore be written as
\begin{equation}
\boldsymbol{\alpha}=
\begin{pmatrix}
\alpha_\perp & i\alpha_H & 0\\
-i\alpha_H & \alpha_\perp & 0\\
0&0&\alpha_\parallel
\end{pmatrix},
\label{eq:alphaMO}
\end{equation}
with $\alpha_H(-\mathbf H_{\rm ext})=-\alpha_H(\mathbf H_{\rm ext})$. The Hall component $\alpha_H$ introduces a helicity-dependent electromagnetic response, thereby coupling the spin angular momentum of thermal photons to radiative heat transport and constitutes the microscopic origin of both the thermal spin Hall effect and its reciprocal inverse effect.

\begin{figure}\centering
\includegraphics[angle=0,scale=0.4]{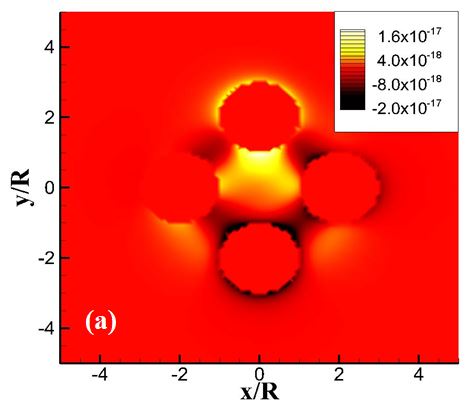}
\includegraphics[angle=0,scale=0.4]{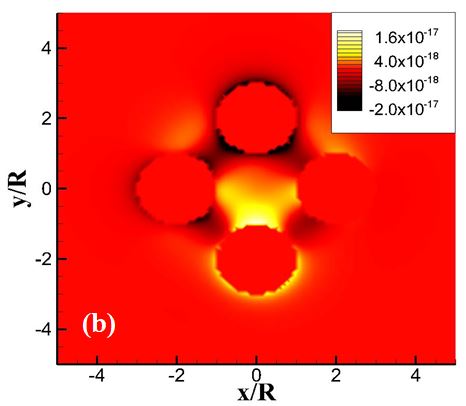}
 \caption{
Spatial distribution of the normal component $S_z$  of spin angular momentum density in the plane $z=0$ around the same $C_4$  network of InSb particles when (a) $\Delta T=100 K$ and (b) $\Delta T=-100 K$  under the action of an external magnetic field applied along $+z$ of magnitude of $H_{ext}=1 T$.
}
\label{Fig3: mapping S vs DT}
\end{figure}

Within the framework of fluctuational electrodynamics
\cite{pbaPRL2011,pbaRMP}, the net power received by the $i^{\rm th}$ particle is
\begin{equation}
\varphi_i
=
\sum_{j\neq i}\varphi_{ji}
+
\varphi_{bi},
\label{eq:net_power}
\end{equation}
where $\varphi_{ji}$ is the power received by particle $i$ from particle $j$, and $\varphi_{bi}$ is the power exchanged with the thermal bath. These powers are given by~\cite{Latella}
\begin{equation}
\varphi_{ji}
=
\int_0^\infty \frac{\rd\omega}{2\pi}
\left[
\Theta_\omega(T_j)\mathcal{T}_{j,i}(\omega,\mathbf H_{\rm ext})
-
\Theta_\omega(T_i)\mathcal{T}_{i,j}(\omega,\mathbf H_{\rm ext})
\right],
\label{eq:pair_flux}
\end{equation}
where the transmission coefficients are calculated using the many-body
Landauer formalism developed in
Refs.~\cite{pbaRMP}
and
\begin{equation}
\varphi_{bi}
=
\int_0^\infty \frac{\rd\omega}{2\pi}
\left[
\Theta_\omega(T_b)-\Theta_\omega(T_i)
\right]
\mathcal T_{b,i}(\omega,\mathbf H_{\rm ext}).
\label{eq:bath_flux}
\end{equation}
Here
\begin{equation}
\Theta_\omega(T)=
\frac{\hbar\omega}{\exp(\hbar\omega/k_B T)-1}
\end{equation}
is the mean energy of a harmonic oscillator at temperature $T$, and $\mathcal T_{j,i}$ is the many-body transmission coefficient from particle $j$ to particle $i$. In a magneto-optical system, Lorentz reciprocity is broken and, in general,
\begin{equation}
\mathcal T_{j,i}(\omega,\mathbf H_{\rm ext})
\neq
\mathcal T_{i,j}(\omega,\mathbf H_{\rm ext}).
\end{equation}
However, microscopic reversibility imposes the Onsager-Casimir symmetry
\begin{equation}
\mathcal T_{j,i}(\omega,\mathbf H_{\rm ext})
=
\mathcal T_{i,j}(\omega,-\mathbf H_{\rm ext}).
\label{eq:microreversibility}
\end{equation}

The spin angular momentum of thermal radiation has recently emerged as
a fundamental quantity in nanoscale radiative heat transfer and
nonreciprocal photonics~\cite{Ott1,Jacob1,Fan2,Jacob2}.The local spin angular momentum (SAM) density of the thermal field is defined as
\begin{equation}
\mathbf S_\omega(\mathbf r)
=
\mathbf S^E_\omega(\mathbf r)
+
\mathbf S^H_\omega(\mathbf r),
\label{eq:spin_density}
\end{equation}
with
\begin{equation}
\mathbf S^E_\omega(\mathbf r)
=
\frac{\varepsilon_0}{2\omega}
\Im
\left[
\left\langle
\mathbf E^*(\mathbf r,\omega)
\times
\mathbf E(\mathbf r,\omega)
\right\rangle
\right],
\end{equation}
and
\begin{equation}
\mathbf S^H_\omega(\mathbf r)
=
\frac{\mu_0}{2\omega}
\Im
\left[
\left\langle
\mathbf H^*(\mathbf r,\omega)
\times
\mathbf H(\mathbf r,\omega)
\right\rangle
\right].
\end{equation}
The total SAM density is then
\begin{equation}
\mathbf S(\mathbf r)
=
\int_0^\infty
\frac{\rd\omega}{2\pi}
\mathbf S_\omega(\mathbf r).
\label{eq:total_spin}
\end{equation}
In the present geometry, the relevant component is the out-of-plane spin angular momentum $S_z$. Indeed, since the external magnetic field is applied along the $z$ axis, it lifts the degeneracy between the two circular polarizations about this axis, giving rise to a finite photon spin polarization parallel to $\mathbf H_{\rm ext}$, whereas the in-plane components vanish by symmetry. The thermal spin Hall response is characterized by the spin contrast between the upper and lower parts of the square,
\begin{equation}
\Delta S_z
=
S_z(\mathbf r_4)-S_z(\mathbf r_3).
\label{eq:spin_contrast}
\end{equation}
Equivalently, the lateral spin-angular-momentum gradient may be estimated as
\begin{equation}
\nabla_y S_z
\simeq
\frac{\Delta S_z}{a}.
\label{eq:spin_gradient}
\end{equation}
Figure~2 summarizes the characteristic signatures of the STHE. The transverse spin accumulation varies linearly with the applied temperature bias, confirming the linear-response regime, and changes sign upon reversal of the temperature bias. It also vanishes in the reciprocal limit $H_{\rm ext}=0$. Indeed, when the external magnetic field is switched off, the Hall component of the polarizability vanishes, $\alpha_H=0$, Lorentz reciprocity is restored, and the $C_4$ symmetry imposes
\begin{equation}
\Delta S_z(H_{\rm ext}=0)=S_z(\mathbf r_4)-S_z(\mathbf r_3)=0.
\end{equation}
The apparent nonzero value close to $H_{\rm ext}=0$ in Figs.~2(b) and 2(c) is not a physical zero-field spin accumulation. In Fig.~2(b), $|\Delta S_z|$ is displayed on a logarithmic scale, on which an exact zero cannot be represented. Moreover, the magnetic-field discretization used to generate the curves is not sufficiently fine in the immediate vicinity of $H_{\rm ext}=0$; the plotted points closest to the origin therefore correspond to small but finite fields. The same finite-field discretization explains the apparent behavior close to the origin in Fig.~2(c). Consequently,
\begin{equation}
\Delta S_z(0,\Delta T)=0,
\qquad
\lim_{H_{\rm ext}\rightarrow0}\Delta S_z(H_{\rm ext},\Delta T)=0.
\end{equation}
A nonzero value of $\Delta S_z$ at finite $H_{\rm ext}$, induced by the longitudinal heat current $J_x^Q$, is the signature of the STHE.

In the linear-response regime, this effect may be written phenomenologically as
\begin{equation}
\nabla_y S_z=\chi_{\rm STHE}(\mathbf H_{\rm ext})J_x^Q,
\label{eq:TSHE_coefficient}
\end{equation}
where $\chi_{\rm STHE}$ denotes the thermal spin Hall susceptibility. The microscopic origin of this response is illustrated in Fig.~3. A longitudinal thermal bias generates opposite photon-spin densities on the two sides of the network, resulting in a finite transverse spin accumulation $\Delta S_z$. Reversing the temperature bias reverses the spin texture while preserving its transverse character, providing the photonic analogue of edge spin accumulation in the electronic spin Hall effect. Since the gyrotropic Hall response changes sign under magnetic-field reversal and disappears at zero field, the STHE contribution satisfies
\begin{equation}
\chi_{\rm STHE}(-\mathbf H_{\rm ext})=-\chi_{\rm STHE}(\mathbf H_{\rm ext}),
\end{equation}
so that reversing the magnetic field reverses the transverse spin accumulation.

The reciprocal process is the inverse spin thermal Hall effect, previously introduced in non-reciprocal many-body systems~\cite{pbaPRL2025}. In that case, a longitudinal gradient of photon spin angular momentum drives a transverse radiative heat flux,
\begin{equation}
\nabla_x S_z
\longrightarrow
J_y^Q.
\label{eq:ISTHE_definition}
\end{equation}
We stress that the quantity calculated directly here is the local photon SAM density and its transverse accumulation $\Delta S_z$, not a conserved spin flux. In a local transport description, electromagnetic spin obeys a balance equation
\begin{equation}
\partial_t S_z+\boldsymbol{\nabla}\!\cdot\mathbf J^S=\tau_z,
\end{equation}
where $\mathbf J^S$ is the spin-angular-momentum current and $\tau_z$ accounts for angular-momentum exchange with matter through absorption, magneto-optical interactions, and optical torque. Thus, strict conservation of the photonic spin sector alone is neither assumed nor required.

The two effects can be described in a common linear-response framework. Introducing the thermodynamic forces
\begin{equation}
X_T=-\frac{\nabla_x T}{T},
\qquad
X_S=-\nabla_x S_z,
\end{equation}
and the transverse fluxes $J_y^Q$ and $J_y^S$, one writes
\begin{equation}
\begin{pmatrix}
J_y^Q\\
J_y^S
\end{pmatrix}
=
\begin{pmatrix}
G_{TT}(\mathbf H_{\rm ext}) & G_{TS}(\mathbf H_{\rm ext})\\
G_{ST}(\mathbf H_{\rm ext}) & G_{SS}(\mathbf H_{\rm ext})
\end{pmatrix}
\begin{pmatrix}
X_T\\
X_S
\end{pmatrix}.
\label{eq:linear_response}
\end{equation}

\begin{figure}\centering
\includegraphics[angle=0,scale=0.4]{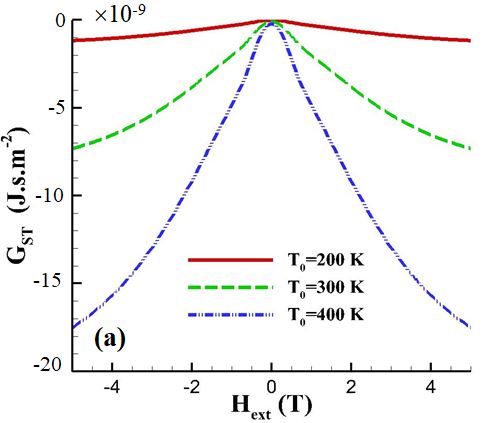}
\includegraphics[angle=0,scale=0.4]{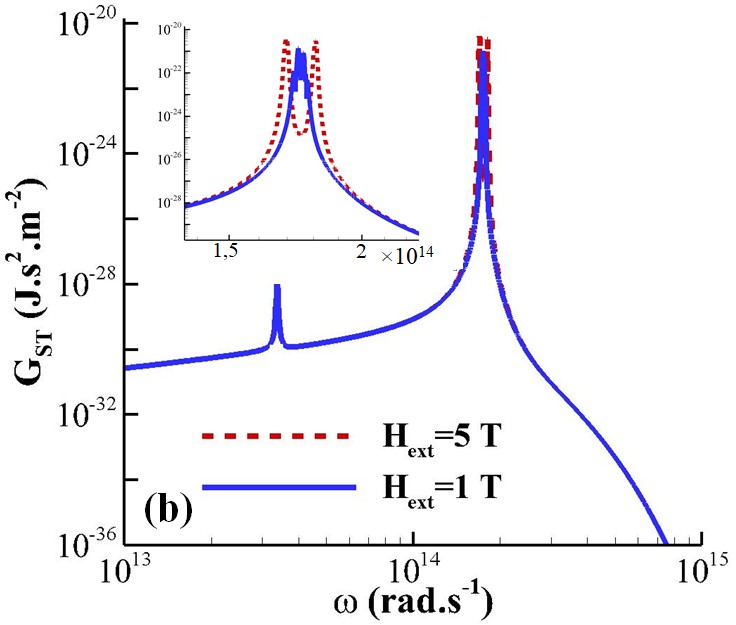}
 \caption{
(a) Thermal spin accumulation coefficient $G_{ST}$ as a function of the external magnetic field $H_{\rm ext}$ applied along the $z$ axis for different equilibrium temperatures $T_0$. (b) Spectral thermal spin accumulation coefficient $G_{ST}(\omega)$ at $T_0=300\,\mathrm{K}$ for $H_{\rm ext}=1\,\mathrm{T}$ and $H_{\rm ext}=5\,\mathrm{T}$. The inset shows a magnified view of the spectral region where the hybridization of resonant modes occurs.
}
\label{Fig4: mapping S vs DT}
\end{figure}

The coefficient $G_{ST}$ describes the direct thermal spin Hall effect, namely the generation of a transverse spin accumulation by a longitudinal thermal driving force or radiative heat current. Conversely, $G_{TS}$characterizes the inverse thermal spin Hall effect, in which a longitudinal spin-angular-momentum gradient induces a transverse radiative heat current.
The coupling between heat and spin transport naturally falls within the
framework of irreversible thermodynamics originally developed by
Onsager~\cite{Onsager1931b} and generalized by
Casimir~\cite{Casimir1945} to systems with broken time-reversal
symmetry.
As a consequence of microscopic reversibility, the Onsager--Casimir
reciprocity relations state that the linear transport coefficients satisfy for two coupled fluxes (i) and (j)
\begin{equation}
G_{ij}(\mathbf H_{\rm ext})=
\epsilon_i\epsilon_j
G_{ji}(-\mathbf H_{\rm ext}),
\label{eq:onsager_general}
\end{equation}
where $\epsilon_i=\pm1$ denotes the parity of the corresponding flux under time reversal. Since both heat and spin currents are odd under time reversal, one obtains
\begin{equation}
G_{TS}(\mathbf H_{\rm ext})=
G_{ST}(-\mathbf H_{\rm ext}),
\label{eq:onsager_TSHE_ISTHE}
\end{equation}
establishing that the thermal spin Hall effect and the inverse thermal spin Hall effect form an Onsager-Casimir reciprocal pair.

The transport coefficients are not merely phenomenological parameters but can be related to the underlying fluctuational-electrodynamic description. A practical expression for $G_{ST}$ is obtained by linearizing the transverse spin accumulation with respect to a small temperature bias $\Delta T=T_1-T_2$ around the equilibrium temperature $T_0$. Using the thermal driving force  $X_T=-\Delta T/(T_0a)$, one get
\begin{equation}
G_{ST}=
\left.
\frac{\partial \Delta S_z}{\partial X_T}
\right|_{T_0}
=-T_0a
\left.
\frac{\partial \Delta S_z}{\partial \Delta T}
\right|_{\Delta T=0}.
\end{equation}
From the expression of the spin angular momentum (SAM) density in terms
of the electric and magnetic field correlation functions, the
spin--thermal response coefficient \(G_{ST}\) is obtained by linearizing
the SAM density with respect to the temperature field around thermal
equilibrium. This yields
\begin{equation}
G_{ST}
=
-T_0a
\int_0^\infty
\frac{\rd\omega}{2\pi}
\left.
\frac{\partial\Theta_\omega}{\partial T}
\right|_{T_0}
\mathcal K_{ST}(\omega,\mathbf H_{\rm ext}),
\label{eq:G_ST}
\end{equation}
where the spectral spin--heat response function is
\begin{equation}
\begin{split}
\mathcal K_{ST}(\omega,\mathbf H_{\rm ext})
=
\frac12
\Big[
&
\mathcal K_1^{S}(\mathbf r_4,\omega)
-
\mathcal K_1^{S}(\mathbf r_3,\omega)
\\
-&
\mathcal K_2^{S}(\mathbf r_4,\omega)
+
\mathcal K_2^{S}(\mathbf r_3,\omega)
\Big],
\end{split}
\end{equation}
with \(\mathcal K_i^{S}=\mathcal K_{i,E}^{S}+\mathcal K_{i,H}^{S}\)
denoting the local spin kernel generated by the fluctuating dipole of
particle \(i\). These kernels are determined by quadratic combinations
of the full many-body electric and magnetic Green tensors together with
the dissipative part of the magneto-optical polarizability tensor. Their
explicit expressions and derivation are provided in the Supplemental
Material~\cite{SupplMat}.The microscopic response coefficient associated with the STHE
follows directly from Eq.~(\ref{eq:G_ST}) and is shown in
Fig.~4(a), while Fig.~4(b) displays its spectral decomposition. As shown in Fig.~4(a), the STHE coefficient vanishes in the reciprocal limit $H_{\rm ext}=0$ and increases in magnitude as the magneto-optical response is activated by the external magnetic field. Its magnitude also increases with equilibrium temperature, indicating enhanced thermal-to-spin conversion. Any apparent finite value at the origin in a logarithmic representation is a graphical/numerical artifact associated with the impossibility of displaying an exact zero on a logarithmic scale and with the finite magnetic-field discretization near $H_{\rm ext}=0$. The spectral
decomposition in Fig.~4(b) reveals that this response is strongly
resonant: a weak contribution appears close to the optical-phonon
band of InSb, whereas the integrated response is dominated by a
narrow high-frequency resonance associated with the free-carrier
dipolar mode of the nanoparticles. Increasing the magnetic field
substantially enhances this dominant resonant contribution through modes hybridization.
The symmetry relation~(\ref{eq:onsager_TSHE_ISTHE}) then provides the reciprocal coefficient $G_{TS}$ directly through magnetic-field reversal, without requiring an independent microscopic calculation. Consequently, once the direct thermal spin Hall response has been evaluated within fluctuational electrodynamics, the complete spin-heat transport matrix is fully determined by microscopic reversibility.

For the four-particle square, the central prediction is that the transverse spin accumulation is proportional to the longitudinal heat current,
\begin{equation}
\Delta S_z
=
\chi_{\rm TSHE}(\mathbf H_{\rm ext})\,J_Q^x,
\end{equation}
where the thermal spin Hall susceptibility changes sign under magnetic-field reversal and vanishes at $H_{\rm ext}=0$. Since the longitudinal heat current satisfies $J_Q^x\propto \Delta T$ in the linear-response regime, the spin accumulation is linear in the applied temperature bias,
\begin{equation}
\Delta S_z
=
\chi_{\rm TSHE}(\mathbf H_{\rm ext})\,\Delta T.
\end{equation}
More generally, the transverse spin accumulation obeys the following symmetry relations
\begin{equation}
\begin{aligned}
\Delta S_z(\mathbf H_{\rm ext},\Delta T)
&=-\Delta S_z(-\mathbf H_{\rm ext},\Delta T)\\
&=-\Delta S_z(\mathbf H_{\rm ext},-\Delta T).
\end{aligned}
\label{eq:symmetry_spin_accumulation}
\end{equation}
These relations provide stringent numerical and experimental signatures of the thermal spin Hall effect: the transverse spin accumulation is odd in both the magnetic field and the applied temperature bias and vanishes at $H_{\rm ext}=0$. The present effect is the reciprocal counterpart of the inverse spin thermal Hall effect. Together, the two phenomena form an Onsager-Casimir pair linked by a unique spin-heat coupling coefficient, establishing the thermodynamic framework of spin-resolved radiative transport and laying the foundations of photon spin caloritronics, with potential applications in spin-controlled thermal radiation and nonreciprocal thermal devices.

Combined with the second law of thermodynamics, the
Onsager-Casimir reciprocity relations
\cite{Casimir1945}
impose strong constraints on the transport matrix governing
spin-resolved radiative transport. In the linear-response regime, the entropy production
rate reads
\begin{equation}
\dot{\Sigma}=J_QX_T+J_SX_S\ge0.
\end{equation}
The positivity of $\dot{\Sigma}$ requires the
symmetric part of the transport matrix to be positive semidefinite,
leading to the bound
\begin{equation}
\left(\frac{G_{TS}+G_{ST}}{2}\right)^2
\le
G_{TT}G_{SS},
\label{eq:bound}
\end{equation}
which limits the strength of the spin--heat coupling and, consequently,
the maximum efficiency of the thermal spin Hall conversion. Therefore, the relative strength of the spin--heat coupling is naturally characterized by the dimensionless parameter
\begin{equation}
\kappa=
\frac{\left(G_{TS}+G_{ST}\right)^2}
{4G_{TT}G_{SS}},
\end{equation}
which compares the spin--heat conversion coefficients to the geometric mean of the direct heat and spin conductances, thereby providing a normalized measure of the coupling between the two transport channels. As a direct consequence of Eq.~(\ref{eq:bound}), it satisfies
\begin{equation}
0\le\kappa\le1.
\end{equation}
By analogy with thermoelectricity, we  introduce the thermal-spin figure of merit,
\begin{equation}
Z_{\rm TS}=
\frac{\kappa}{1-\kappa},
\end{equation}
which quantifies the efficiency of the mutual conversion between heat transport and photon spin. This quantity plays for photon spin caloritronics an analogous role as the thermoelectric figure of merit $ZT$ in thermoelectricity, providing a universal metric for quantifying the efficiency of spin-heat conversion in radiative systems.The limits $Z_{\rm TS}\ll1$ and
$Z_{\rm TS}\rightarrow\infty$
correspond to weak and maximally coupled spin--heat transport,
respectively. In the latter limit, the spin--heat conversion reaches the
maximum value permitted by the second law of thermodynamics.

The central conceptual advance is therefore not the formal existence of an Onsager-reciprocal coefficient, but the independent microscopic demonstration that thermal driving can generate photon spin. Together with the previously demonstrated spin-to-heat conversion, this establishes the bidirectional conversion heat to photon\ spin and provides the physical basis for radiative spin caloritronics.

In conclusion, we have predicted the spin thermal Hall effect in nonreciprocal photonic many-body systems and shown that this phenomenon is the Onsager--Casimir reciprocal counterpart of the ISTHE, thereby establishing a unified thermodynamic framework for coupled heat and photon-spin transport. These results identify photon spin as a genuine transport variable in fluctuational electrodynamics and complete the thermodynamic description of spin-resolved radiative transport. More broadly, they provide unprecedented opportunities for controlling thermal radiation through the spin degree of freedom of light and establish the conceptual foundations of photon spin caloritronics and spin-controlled radiative thermal devices.

\begin{acknowledgments}
\end{acknowledgments}

\end{document}